\documentclass[a4paper,oneside,english,oneside,reqno]{amsart}
\usepackage[T1]{fontenc}
\usepackage[latin9]{inputenc}
\usepackage{fancyhdr}
\usepackage{units}
\usepackage{amsthm}
\usepackage{amstext}
\usepackage{graphicx}
\usepackage{setspace}
\usepackage{amssymb}

\makeatletter
\numberwithin{equation}{section} 
\numberwithin{figure}{section} 
\theoremstyle{plain}

\makeatletter
\numberwithin{equation}{section} 
\numberwithin{figure}{section} 

\usepackage{esint}

\makeatletter


\makeatother

\makeatother

\makeatother

\usepackage{babel}

\begin{document}

\title{Learning, complexity and information density}

\author{Joel Ratsaby}

\maketitle
\begin{singlespace}
\begin{center}
Department of Electrical and Electronics Engineering, Ariel University
Center, Ariel 40700, ISRAEL
\par\end{center}

\begin{center}
\texttt{ratsaby@ariel.ac.il} 
\par\end{center}
\end{singlespace}

\begin{abstract}
What is the relationship between the complexity of a learner and the
randomness of his mistakes ? This question was posed in \cite{rat0903}
who showed that the more complex the learner the higher the possibility
that his mistakes deviate from a true random sequence. In the current
paper we report on an empirical investigation of this problem. We
investigate two characteristics of randomness, the stochastic and
algorithmic complexity of the binary sequence of mistakes. A learner
with a Markov model of order $k$ is trained on a finite binary sequence
produced by a Markov source of order $k^{*}$ and is tested on a different
random sequence. As a measure of learner's complexity we define a
quantity called the \emph{sysRatio}, denoted by $\rho$, which is
the ratio between the compressed and uncompressed lengths of the binary
string whose $i^{th}$ bit represents the maximum \emph{a posteriori}
decision made at state $i$ of the learner's model. The quantity $\rho$
is a measure of information density. The main result of the paper
shows that this ratio is crucial in answering the above posed question.
The result indicates that there is a critical threshold $\rho^{*}$
such that when $\rho\leq\rho^{*}$ the sequence of mistakes possesses
the following features: (1)\emph{ }low divergence $\Delta$ from a
random sequence, (2) low variance in algorithmic complexity. When
$\rho>\rho^{*}$, the characteristics of the mistake sequence changes
sharply towards a\emph{ }high\emph{ $\Delta$} and high variance in
algorithmic complexity. 
\end{abstract}

\section{\label{sec1}Overview}

In computer science, the notion of computational complexity serves
as a measure of how difficult it is to compute a solution for a given
problem. Computations take time and complexity here means the time
rate of growth to solve the problem. Another related kind of complexity
measure (studied in theoretical computer science) is the so-called
algorithmic (or Kolmogorov) complexity which measures how long a computer
program (on some generic computational machine) needs to be in order
that it produces a complete description of an object. Interestingly,
the theory says that if we consider as an object a system that can
process input information (available as a binary sequence of high
entropy) and which produces another sequence as an output then the
amount of randomness in the output sequence is inversely proportional
to the algorithmic complexity of the system.

This has been traditionally studied in the context of algorithmic
randomness (see \cite{LaurentBienvenu07} and references within) and
it has been only until recently unknown whether such a relationship
between complexity and randomness exists for more general systems,
for instance, those governed by physical laws. In \cite{Ratsaby_entropy}
the complexity of a general static system (for instance, a physical
solid) is modeled algorithmically, i.e., by its description length.
Using the model it is proposed that the stability of a static system
(from the physical perspective) is related to its level of algorithmic
complexity. This is explained by the relationship between the complexity
of a system and its ability to 'distort' the randomness in its environment.
The first proof of this concept appeared in a recent paper \cite{ratChasAIPR09}
where it was shown that this inverse relationship between system complexity
and randomness exists also in a physical system. The particular system
investigated consisted of a one-dimensional vibrating solid-beam to
which a random sequence of external input forces is applied. 

The current paper is yet another proof of concept of the model of
\cite{Ratsaby_entropy}. We proceed along the line of \cite{ratChasAIPR09}
but instead of considering a physical system (the static solid with
input force sequence) we consider a \emph{decision} system and study
its influence on a random binary data sequence on which prediction
decisions are made. The decision system is based on the maximum \emph{a
posteriori} probability decision where probabilities are defined by
a statistical parametric model which is estimated from data. The learner
of this model is a computer program that trains from a given random
data sequence and then produces a decision rule by which it is able
to predict (or decide) the value of the next bit in future (yet unseen)
random binary sequences. 

While this paper is in the realm of machine-learning we are not proposing
a new algorithm nor are we interested in the performance of the learner.
But rather, our interest is in displaying a learning (and decision)
system from the perspective of static system complexity and its influence
on random inputs \cite{Ratsaby_entropy}.

\section{Introduction}

Let $X^{(n)}=X_{1},\ldots,X_{n}$ be a sequence of binary random variables
drawn according to some unknown joint probability distribution $\mathbb{P}\left(X^{(n)}\right)$.
Consider the problem of learning to predict the next bit in a binary
sequence drawn according to $\mathbb{P}$. For training, the learner
is given a finite sequence $x^{(m)}$ of bits $x_{t}\in\left\{ 0,1\right\} ,$
$1\leq t\leq m$, drawn according to $\mathbb{P}$ and estimates a
model $\mathcal{M}$ that can be used to predict the next bit of a
partially observed sequence. After training, the learner is tested
on another sequence $x^{(n)}$ drawn according to the same unknown
distribution $\mathbb{P}$. Using $\mathcal{M}$ he produces the bit
$y_{t}$ as a prediction for $x_{t}$ , $1\leq t\leq n$. Denote by
$\xi^{(n)}$ the corresponding binary sequence of mistakes where $\xi_{t}=1$
if $y_{t}\neq x_{t}$ and is $0$ otherwise. In \cite{rat0903} the
following question was posed: how random is $\xi^{(n)}$ ?

It is clear that the sequence of mistakes should be random since the
test sequence $x^{(n)}$ is random. It may also be that because the
learner is using a model of a finite structure (or a finite description-length)
that it may somehow introduce dependencies and cause $\xi^{(n)}$
to be less random than $x^{(n)}$. And yet by another intuition, perhaps
the fact that the learner is of a finite complexity limits its ability
to 'deform' (or distort) randomness of $x^{(n)}$ ? These are all
valid initial guesses that relate to this main question. We note that
our basis for saying that $\mathcal{M}$ has a finite structure stems
from it being an element of some regular hypothesis class, for instance,
having a finite VC-dimension as is often the case in a learning setting
(see for instance structural risk minimization of \cite{Vapnik1998}).
In the current paper, we are not interested in the learner's performance
(as modeled for instance by Valiant's PAC framework \cite{Valiant84,EncyAlgo08})
but instead we take a black-box view of a learner and ask how much
influence does he has on the stochastic properties of the errors.
We view the learner as an entity that 'interferes' with the randomness
that is inherent in the sequence to be predicted and through his predictions
creates a sequence of mistakes that has a different stochastic character.
This view in a broader sense is taken in \cite{Ratsaby_entropy} and
is shown (empirically) in \cite{ratChasAIPR09} to explain how static
structures may 'deform' random external forces.

The question raised above was answered in \cite{rat0903} for a particular
learning setting where the teacher uses a probability distribution
$\mathbb{P}$ based on a Markov model with a certain complexity. The
learner has access to a hypothesis class of Boolean decision rules
that are based on Markov models. Hence, learning amounts to the estimation
of parameters of a finite-order Markov model (see for instance \cite{Kemeny76,Medhi94}).
The answer shows theoretically that the random characteristics of
the subsequence of mistakes corresponding to the $0$-predictions
of a learner changes in accordance with the complexity of the learner's
decision rule's complexity. The more complex the rule the higher the
possibility of 'distortion' of randomness, i.e., the farther away
it is from being truly-random. 

In the current paper we take an experimental approach to answering
the above question. As in \cite{rat0903} we focus on Markov source
and a Markov learner whose orders may differ. In the next section
we describe the setup.

\section{\label{sec:Experimentl-setup}Experimentl setup}

The learning problem consists of predicting the next bit in a given
sequence generated by a Markov chain (model) $\mathcal{M^{*}}$ of
order $k^{*}$. There are $2^{k^{*}}$ states in the model each represented
by a word of $k^{*}$ bits. During a learning problem, the source's
model is fixed. A learner, unaware of the source's model, has a Markov
model of order $k$. We denote by $p(1|i)$ the probability of transiting
from state $i$ whose binary $k$-word is $b_{i}=[b_{i}(1),\ldots,b_{i}(k)]$
to the state whose word is $[b_{i}(2),\ldots,b_{i}(k),1]$. Given
a random sequence of length $m$ generated by the source the learner
estimates its own model's parameters $p(1|i)$ by $\hat{p}(1|i)$,
$1\leq i\leq2^{k}$, which is the frequency of the event {}``$b_{i}$
is followed by a $1$'' in the training sequence. We denote by $\hat{\mathcal{M}}$
the learnt model with parameters $\hat{p}(1|i)$, $1\leq i\leq2^{k}$.
We denote by $p^{*}(1|i)$ the transition probability from state $i$
of the source model, $1\leq i\leq2^{k}$.

A simulation run is characterized by the parameters, $k$ and $m$.
It consists of a training and testing phases. In the training phase
we show the learner a binary sequence of length $m$ and he estimates
the transition probabilities. In the testing phase we show the learner
another random sequence (generated by the same source) of length $n$
and test the learner's predictions on it. For each bit in the test
sequence we record whether the learner has made a mistake. When a
mistake occurs we indicate this by a $1$ and when there is no mistake
we write a $0$. The resulting sequence of length $n$ is the generalization
mistake sequence $\xi^{(n)}$. We denote by $\xi_{0}^{(n)}$ the binary
subsequence of $\xi^{(n)}$ that corresponds to the mistakes that
occured only when the learner predicted a $0$.

For a fixed $k$ denote by $N_{k,m}$ the number of runs with a learner
of order $k$ and training sample of size $m$. The experimental setup
consists of $N_{k,m}=10$ runs with $1\leq k\leq10$, $m\in\left\{ 100,200,\ldots,10000\right\} $
with a total of $100\cdot10\cdot N_{k,m}=10000$ runs. The testing
sequence is of length $n=1000$. Each run results in a file called
\emph{system} which contains a binary vector $d$ whose $i^{th}$
bit represents the maximum \emph{a posteriori} decision made at state
$i$ of the learner's model, i.e., \begin{equation}
d_{i}=\left\{ \begin{array}{cc}
1 & \text{if \, }\hat{p}(1|i)>\nicefrac{1}{2}\\
0 & otherwise\end{array}\right.\label{eq:zi}\end{equation}
for $1\leq i\leq2^{k}$. Let us denote by $\alpha_{i}=P(\hat{p}(1|i)>\nicefrac{1}{2})$,
thus $d_{i}$ are Bernouli random variables with parameters $\alpha_{i}$,
$1\leq i\leq2^{k}$. The learner's system is its decision rule at
every possible state.

Another file generated is the \emph{errorT0} which contains the mistake
subsequence $\xi_{0}^{(n)}$. At the end of each run we measure the
lengths of the \emph{system} file and its compressed length where
compression is obtained via the Gzip algorithm (a variant of \cite{LZ77})
and compute the \emph{sysRatio} (denoted as $\rho)$ which is the
ratio of the compressed to uncompressed length of the system file.
Note that $\rho$ is a measure of information density since it captures
the number of bits of useful information (useful for describing the
system) there are per bit of representation (in the uncompressed file). 

We do similarly for the mistake-subsequence $\xi_{0}^{(n)}$ obtaining
the length $\ell_{0}$ of the compressed file that contains $\xi_{0}^{(n)}$
(henceforth referred to as the estimated algorithmic complexity of
$\xi_{0}^{(n)}$ since it is an approximation of the Kolmogorov complexity
of $\xi_{0}^{(n)}$, see \cite{ratChasAIPR09}). We measure the KL-divergence
$\Delta_{0}$ between the probability distribution $P(w|\hat{p})$
of binary words $w$ of length $4$ and the empirical probability
distribution $\hat{P}_{m}(w)$ as measured from the mistake subsequence
$\xi_{0}^{(n)}$. Note, $P(w|\hat{p})$ is defined according to the
Bernouli model with parameter $\hat{p}$, that is, $P(w|\hat{p})=\hat{p}^{i}(1-\hat{p})^{4-i}$
for a word $w$ with $i$ ones, where $\hat{p}$ is the frequency
of ones in the subsequence $\xi_{0}^{(n)}$. The distribution $\hat{P}_{m}(w)$
equals the frequency of a word $w$ in $\xi_{0}^{(n)}$. Hence $\Delta_{0}$
reflects by how much $\xi_{0}^{(n)}$ deviates from being random according
to a Bernoulli sequence.

\section{Results}

We are interested in the determining the following relationships:
(1) the system ratio $\rho$ versus the learner's model order $k$,
(2) the estimated algorithmic complexity $\ell_{0}$ of the subsequence
$\xi_{0}^{(n)}$ versus the $\rho$, and (3) the deviation $\Delta_{0}$
versus $\rho$. 

We choose four different levels of learning problems, controlled by
the order of the source model $k^{*}=3$, $4$, $5$, $6$. For each
problem we choose for the source model a transition matrix of probabilities
$p^{*}(1|i)=1-p$, $p^{*}(0|i)=p$, where for some of the states $i$
we set $p=0.3$ and for others $p=0.7$, $1\leq i\leq2^{k^{*}}$.
Thus the Bayes optimal error is $0.3$. To ensure that the problem
is sufficiently challenging we set the first half of the states (those
ranging from the $k^{*}$-dimensional vector $00\ldots0$ to $011\ldots1$)
to have $p=0.3$ and the second half ($10\ldots0$ to $11\ldots1$)
to have $p=0.7$. This ensures that a Markov model of order $k<k^{*}$
cannot approximate the true transition probabilities well, i.e., the
infinite-sample limit estimate based on a Markov model of order $k$
which is smaller than $k^{*}$ will still be $\hat{p}(1|i)=0.5$,
$1\leq i\leq2^{k}$. But for a Markov model of order $k\geq k^{*}$
the infinite-sample size estimates will converge to the true values
of $p$ or $1-p$. 

\begin{figure}
\begin{centering}
\includegraphics[bb=0bp 580bp 564bp 800bp,scale=0.8]{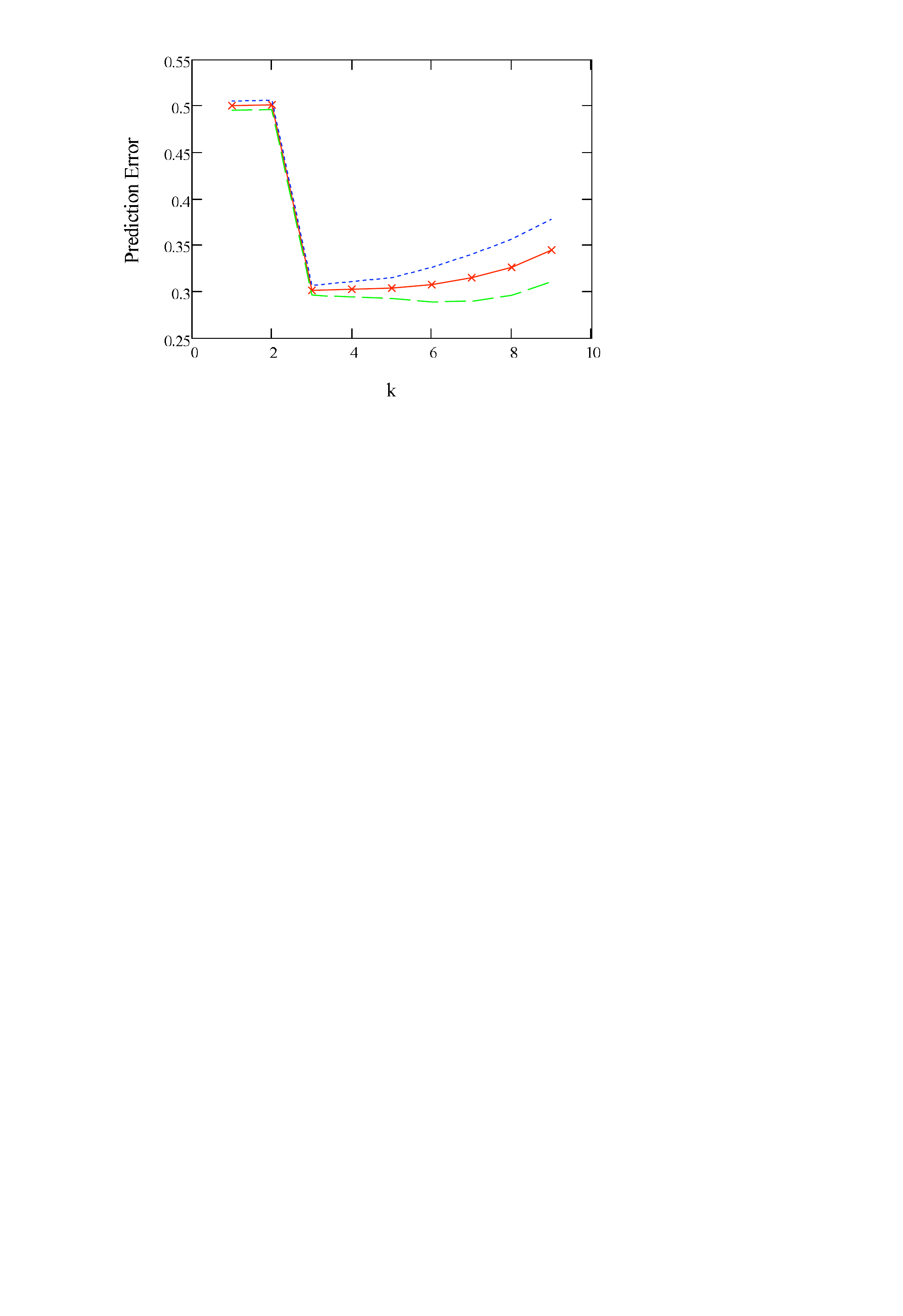}
\par\end{centering}

\caption{\label{fig:generalization-error-versus}generalization error versus
$k$ for $k^{*}=3$ }

\end{figure}

Before we start to investigate the three relationships stated above
we perform a sanity check to see how the prediction generalization
error (for any of the two prediction types, not just when predicting
a zero) varies with respect to the model complexity $k$. Figure \ref{fig:generalization-error-versus}
displays this relationship for a learning problem with $k^{*}=3$.
The curve (with $\mathsf{x}$) is the mean error over all learning
runs of a fixed $k$ value, the upper and lower curves are the standard
deviation above and below the mean, respectively. As seen, when the
learner's model order $k$ is smaller than $k^{*}$ his generalization
error stays at the maximum level of $0.5$. At $k=k^{*}$ there is
a drop to an error close to the Bayes error of $0.3$ Then as $k$
increases beyond $k^{*}$ the mean (as well as the standard deviation)
of the generalization error start to increase. This is due to overfitting
of the model to the training data and also because the variance of
the error estimate increases with $k$ due to the fact that the maximum
sample size of any run is fixed at $m=10000$ and is not increasing
with respect to $k$.

\begin{figure}
\begin{raggedright}
\includegraphics[bb=80bp 330bp 582bp 600bp,scale=0.6]{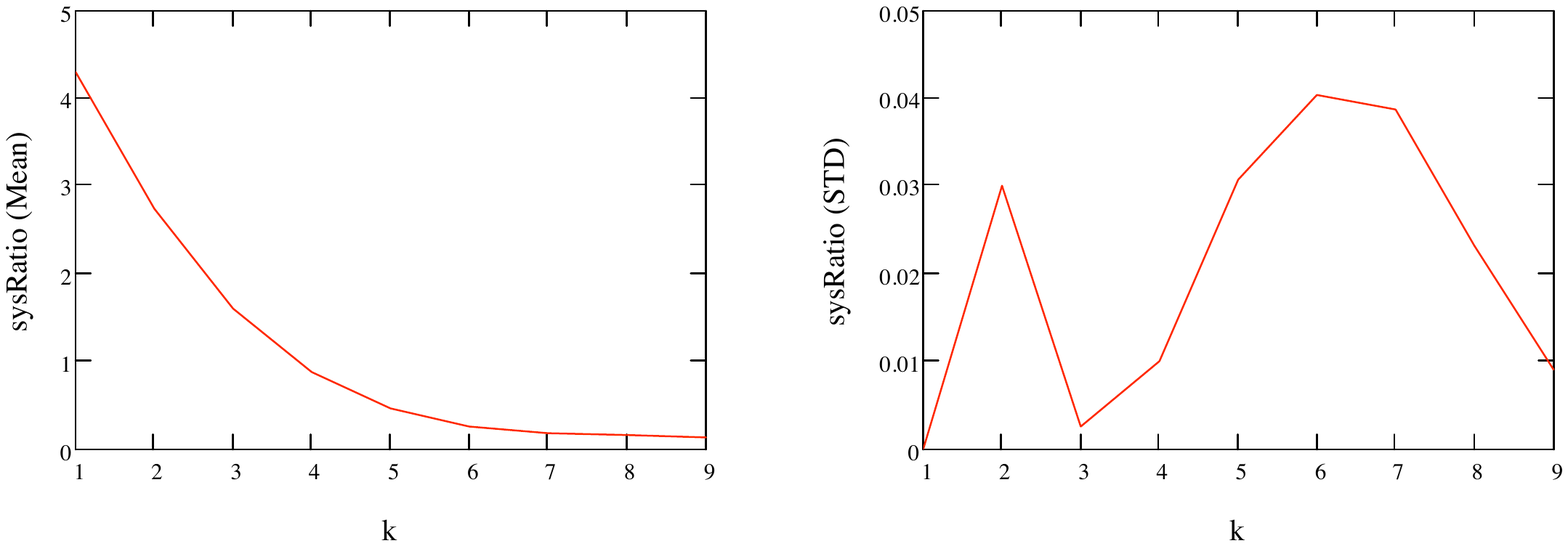}
\par\end{raggedright}

\caption{\label{fig:SysRatio--versus}The mean of SysRatio versus $k$ (left),
the standard deviation of SysRatio versus $k$ (right)}

\end{figure}

We now proceed to describe the first result which concerns  the relationship
between the sysRatio $\rho$ and $k$. Figure \ref{fig:SysRatio--versus}
shows the mean and standard deviation of the SysRatio $\rho$ as a
function of $k$. The mean decreases as the learner's model order
$k$ increases. To explain this, first note that the uncompressed
length of the system is always $c\cdot2^{k}$ for some constant $c>0$
since the vector $d$ is of length $2^{k}$ (see section \ref{sec:Experimentl-setup}).
The length of the compressed system file also grows, but at a slower
rate with respect to $k$ and this gives rise to the decrease in $\rho$
with respect to $k$. Why is the rate of the compressed system file
growing more slowly ? 

The reason is that for values of $k<k^{*}$ the learner's model is
incapable (by design of the learning problem) of estimating the Bayes
optimal prediction and the probability of the events {}``$b_{i}$
is followed by a $1$'' is $p(1|i)=\nicefrac{1}{2}$ , $1\leq i\leq2^{k}$.
Thus the average value $\hat{p}(1|i)$ of the indicators of such events
is a Binomial random variable with a distribution symmetric at $\nicefrac{1}{2}$
and hence from (\ref{eq:zi}) the probability $\alpha_{i}$ that $\hat{p}(1|i)>\nicefrac{1}{2}$
equals $\nicefrac{1}{2}$. The components of the random vector $d$
are independent Bernouli random variables with parameter $\alpha_{i}$
when conditioned on the sample size vector $v$ (this is the vector
whose components $v_{i}$ are the number of times that $b_{i}$ appeared
in the training sequence, see \cite{rat0903} for details). Since
in this case $\alpha_{i}=\nicefrac{1}{2}$ then each component has
a maximum entropy $H(d_{i})=-\alpha_{i}\log\alpha_{i}-(1-\alpha_{i})\log(1-\alpha_{i})=\log2=1$
and hence the expected value of the entropy of the vector $d$ (with
respect to the random sample size vector $v$) is maximal and equals
$E_{v}H(d|v)=E_{v}\sum_{i=1}^{2^{k}}H(d_{i}|v_{i})=E_{v}2^{k}=2^{k}.$
Hence the expected compressed length of the system file (which contains
the vector $d$) is large as the expected description length of any
random variable is at least as large as its entropy. 

As $k$ increases beyond $k^{*}$ the model becomes more capable of
estimating the true transition probabilities (recall, these are either
$0.3$ or $0.7$) and the probability $p(1|i)$ of the events {}``$b_{i}$
is followed by a $1$'' get farther away from $\nicefrac{1}{2}$
in the direction of $0.3$ or $0.7$, depending on the particular
state $i$, $1\leq i\leq2^{k}$. Thus the average value $\hat{p}(1|i)$
of the indicators of such events is a Binomial random variable with
an asymmetric distribution with a mean $p(1|i$). Hence from (\ref{eq:zi})
the probability $\alpha_{i}$ that $\hat{p}(1|i)>\nicefrac{1}{2}$
gets either very close to $0$ or $1$ as the training size $m$ increases.
Thus the components of the random vector $d$ tend to be closer to
deterministic. They are still random since the training sequence length
is not increasing with $k$ and the variance of the estimates $\hat{p}(1|i)$
does not converge to zero. Therefore for each of the $2^{k}$ components
of the vector $d$ the entropy is smaller than when $k<k^{*}$. However
as there are exponentially many components $d_{i}$, on the whole,
the entropy of $d$ (and hence the expected compressed length of the
system file) still increase but at a lower rate than when $k<k^{*}$. 

\begin{figure}
\begin{centering}
\includegraphics[scale=0.7]{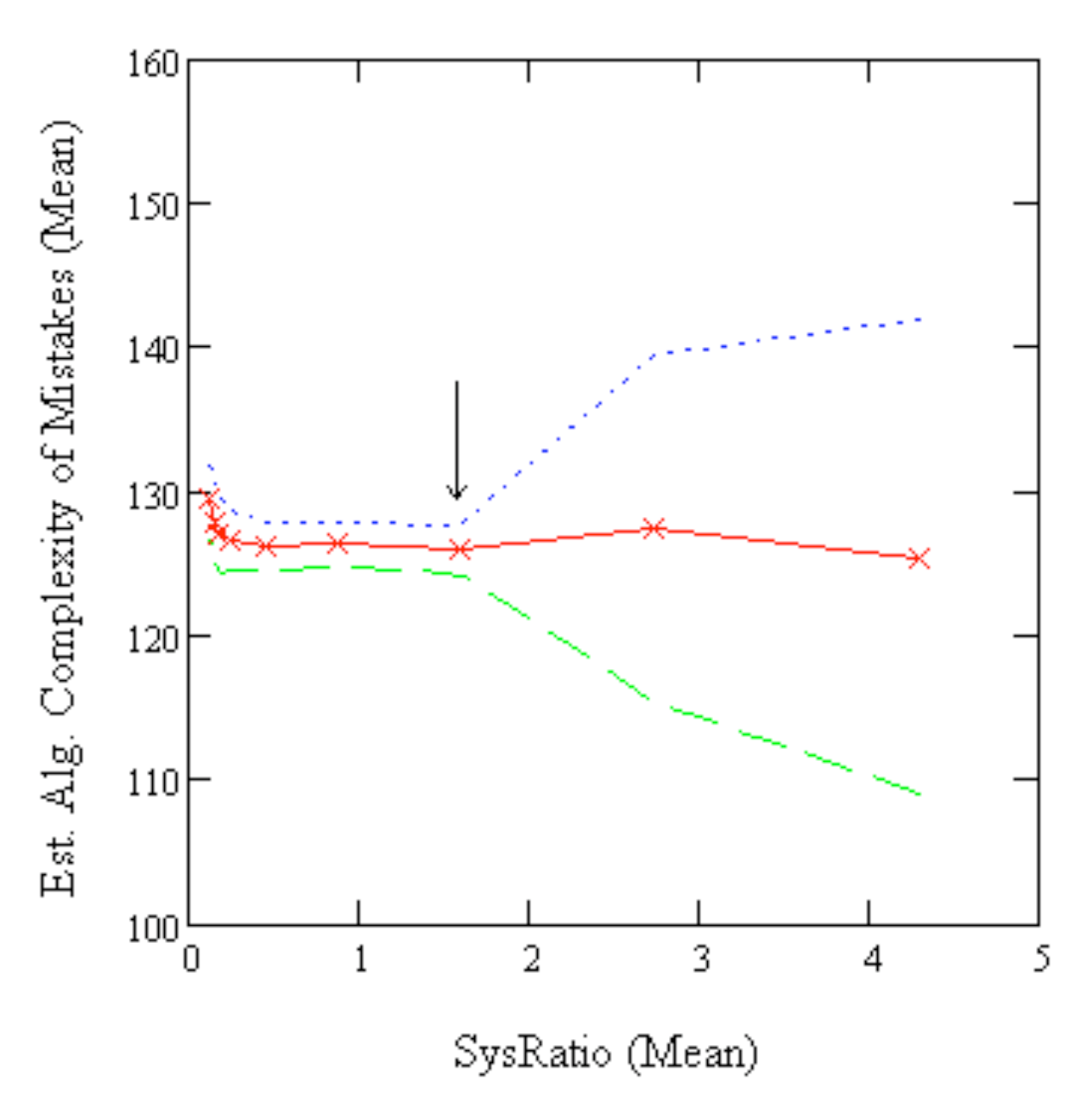}
\par\end{centering}

\caption{\label{fig:Algorithmic-complexity-}Estimated algorithmic complexity
$\ell_{0}$ of the mistake subsequence $\xi_{0}^{(n)}$ versus the
SysRatio $\rho$}

\end{figure}

Next, we discuss the characteristics of the mistake subsequence $\xi_{0}^{(n)}$.
Figure \ref{fig:Algorithmic-complexity-} shows the graph (with $\mathsf{x}$)
of the mean of the estimated algorithmic complexity $\ell_{0}$ of
$\xi_{0}^{(n)}$ versus the mean of the system ratio $\rho$ on the
horizontal axis. The dashed lines are the upper and lower envelopes
of the standard deviation from  the mean. The arrow points at the
value of $\rho^{*}$ that corresponds to $k^{*}$ (the source model
order). As can be seen, for low values of sysRatio the spread $\ell_{0}$
is low. There is a sharp threshold at $\rho^{*}$ where the spread
around the mean value of $\ell_{0}$ increases significantly.

\begin{figure}
\includegraphics[scale=0.55]{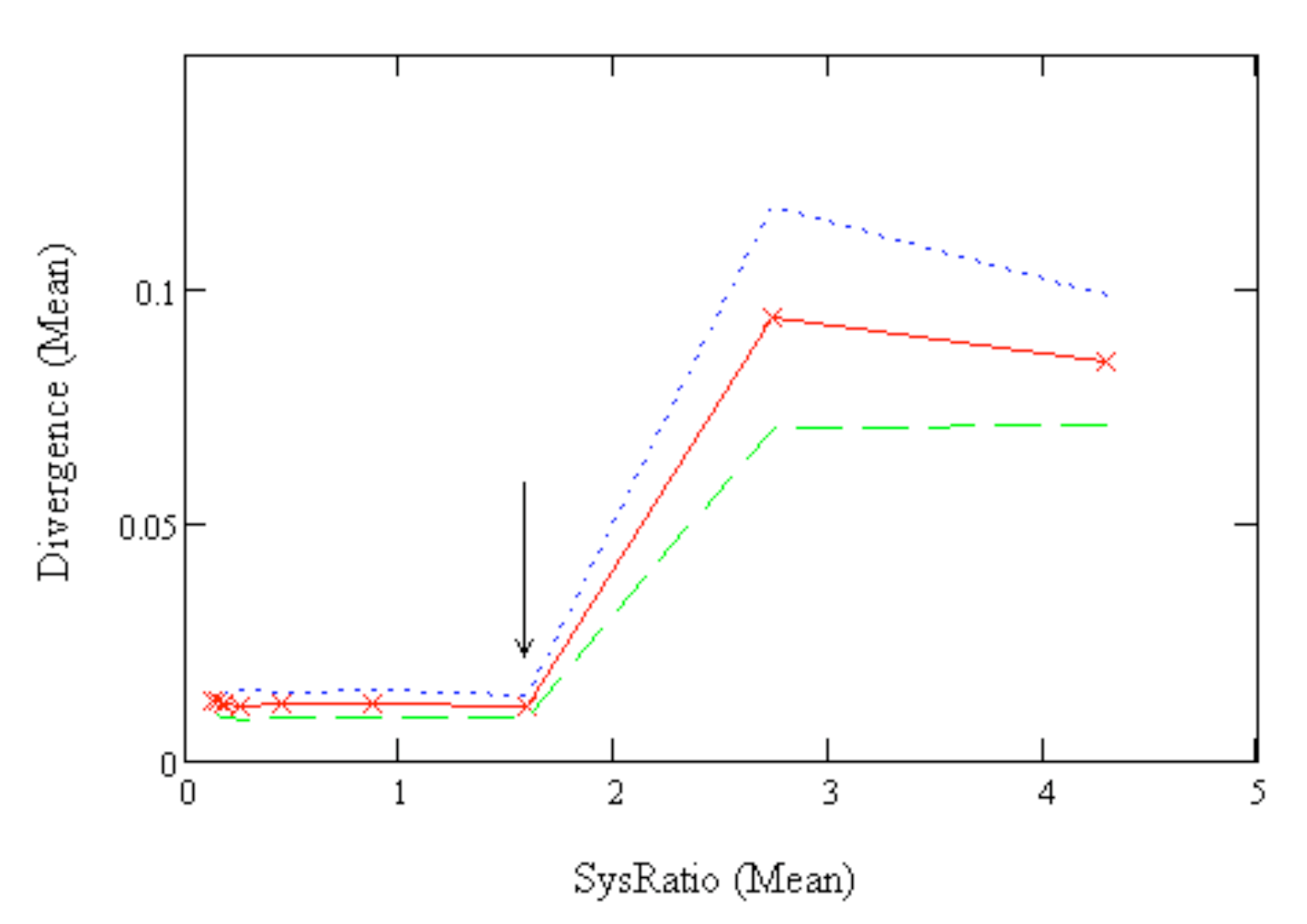}

\caption{\label{fig:Divergence--of}Divergence $\Delta_{0}$ of the mistake
subsequence $\xi_{0}^{(n)}$ versus the SysRatio $\rho$ }

\end{figure}

Next, Figure \ref{fig:Divergence--of} displays the graph (with $\mathsf{x}$)
of the mean of the divergence $\Delta_{0}$ of the mistake subsequence
$\xi_{0}^{(n)}$ versus the mean of the system ratio $\rho$ on the
horizontal axis. The dashed lines are the upper and lower envelopes
of the standard deviation from the mean. The arrow points at the value
of $\rho^{*}$ that corresponds to $k^{*}$ (the source model order).
As can be seen, for low values of sysRatio the spread of $\Delta_{0}$
is low. As the result above for $\ell_{0}$, we see a threshold at
$\rho^{*}$ where the standard deviation around the mean value of
$\Delta_{0}$ increases significantly.

\section{Conclusions}

The paper introduces the notion of sysRatio $\rho$ which is a measure
of information density of the learner's model. It is similar to the
notion of rate of information transmission \cite{CoverThomas91} as
it measures the ratio of the number of useful information bits contained
in a file that describes the learner decision rule per bit of representation
(in the file). The results of this paper depict that this information
density influences the level of randomness of the mistakes made by
a learner. The sysRatio $\rho$ is a proper measure of complexity
of a learner decision rule. It is with respect to $\rho$ that the
characteristics of the random mistake subsequence $\xi_{0}^{(n)}$
follow what the theory \cite{rat0903} predicts. The higher the sysRatio
the more significant the deviation $\Delta_{0}$ of $\xi_{0}^{(n)}$
compared to a pure Bernouli random sequence. In addition, we have
shown that the higher the sysRatio the larger the possible fluctuations
in the algorithmic complexity $\ell_{0}$ of $\xi_{0}^{(n)}$. The
interesting point is the sharp non-linearity in this relationship.
We showed that there is a threshold $\rho^{*}$ at which the spread
in values of $\ell_{0}$ and $\Delta_{0}$ increases and it corresponds
to the point where the learner's model becomes too simple and is incapable
of predicting well.

\bibliographystyle{plain}

\end{document}